\documentclass[amsmath,aps,twocolumn,prl,showpacs,showkeys,groupedaddress]{revtex4}

\usepackage{graphicx}
\usepackage{xspace}
\usepackage{color}
\usepackage{filecontents}

\usepackage{epstopdf}
\newcommand{\gpvec}[1]{\mathbf{#1}}
\newcommand{\qvec}{\gpvec{q}}
\newcommand{\kvec}{\gpvec{k}}
\newcommand{\xvec}{\gpvec{x}}

\newcommand{\imag}{\i}
\newcommand{\plaind}{\mathrm{d}}

\newcommand{\phitilde}{\tilde{\phi}}
\newcommand{\etatilde}{\tilde{\eta}}

\newcommand{\latin}[1]{{\it #1}}
\newcommand{\ie}{\latin{i.e.}\@\xspace}
\newcommand{\eg}{\latin{e.g.}\@\xspace}

\newcommand{\elabel}[1]{\label{eq:#1}}

\newcommand{\Eref}[1]{Eq.~(\ref{eq:#1})}

\newcommand{\flabel}[1]{\label{fig:#1}}

\newcommand{\Fref}[1]{Fig.~\ref{fig:#1}}

\newcommand{ \citecommented}[2]{(#1 \cite{#2})}

\newcommand{\half}{\mathchoice{\frac{1}{2}}{(1/2)}{\frac{1}{2}}{(1/2)}}

\begin{document}
\title{Mounding in Epitaxial Surface Growth}
\author{Edward Sherman}
\email{edward.sherman04@imperial.ac.uk}
\affiliation{Department of Mathematics,
Imperial College London,
180 Queen's Gate,
London SW7 2BZ, United Kingdom}
\author{Gunnar Pruessner}
\email{g.pruessner@imperial.ac.uk}
\affiliation{Department of Mathematics,
Imperial College London,
180 Queen's Gate,
London SW7 2BZ, United Kingdom}

\date{\today}

\begin{abstract}
For the past two decades the Villain-Lai-Das Sarma equation has served
as the theoretical framework for conserved surface growth processes,
such as molecular-beam epitaxy. However some phenomena, such as
mounding, are yet to be fully understood. In the following, we present a
systematic analysis of the full, original Villain-Lai-Das Sarma equation
showing that mound forming terms, which should have been included
initially on symmetry grounds, are generated under renormalisation. A
number of widely studied Langevin equations are recovered as limits or
trivial fixed points of the full theory.
\end{abstract}


\maketitle

Crystal surface growth, in which the interface is driven by the
deposition of new material from a process such as molecular-beam epitaxy
(MBE), has been extensively studied
\cite{barabasistanley,pimpinellicrystals,Krug:1997,KrugMichely2004}.
Although the behaviour of surface growth in the absence of conservation
generically belongs to the Kardar-Parisi-Zhang (KPZ) \cite{kpz}
universality class, frequently the growth condition for MBE allow the
imposition of conservation laws that prohibit KPZ behaviour. Typically
experimental observations of MBE on semiconductor wafers are compared
with numerical simulations run on lattice models, popularly the
Wolf-Villain (WV) \cite{wolfvillain90} and the Das Sarma-Tamborenea (DT)
\cite{DTmodelPRL91} models, and continuum Langevin equations postulated
from physical considerations. Analytically, epitaxial surface growth is
almost exclusively modeled using the Villain-Lai-Das Sarma (VLDS)
\cite{Villain91,LaiDasSarmaPRL91} equation. Although successful, the
theoretical framework seems somewhat incomplete. No clear picture has
emerged over the alleged exactness of scaling relations and the
mechanism of mound formation is unaccounted for. In the following we
explain why mounding is, and ought to be, observed either transiently or
stably in MBE/VLDS models; mound formation arises naturally from a term
in the Langevin equation that has been variously missed, overlooked or
discarded in the literature. A resolution to apparent disagreements and
misunderstandings on the nature of the scaling relations and coupling
renormalisation is offered. A complete theoretical picture for the
behaviour and the different regimes of epitaxial surface growth as
originally envisaged two decades ago is presented, with the full theory
for conserved surface growth via ideal MBE given by:
\begin{multline}\elabel{full_theory}
\partial_{t}\phi(\xvec,t)=
\nu_{2}\nabla^{2}\phi-
\nu_{4}\nabla^{4}\phi+
\lambda_{13}\nabla\Big(\nabla\phi\Big)^{3}\\
\qquad+\tilde{\lambda}_{22}\nabla^{2}\Big(\nabla\phi\Big)^{2}
+\tilde{\kappa}\nabla\left(\nabla\phi\nabla^{2}\phi\right)+\eta(x,t),
\end{multline}
where the field $\phi(\xvec,t)$ is the surface (height) displacement at
$\xvec$ in the co-moving frame atop a $d$ dimensional substrate at time
$t$. Growth is subject to the white noise $\eta$ with the usual
correlator
$\langle\eta(x,t)\eta(x',t')\rangle=2\Gamma^{2}\delta^{d}(x-x')\delta(t-t')$.
Ideal MBE \cite{LaiDasSarmaPRL91} was proposed as ``atomistic stochastic
growth without any bulk defects or surface overhangs driven by atomic
deposition in a chemical-bonding environment where surface relaxation
can occur only through the breaking of bonds.'' This constrains the
surface dynamics to obey mass conservation; the deterministic evolution
being cast as the divergence of some current, thus ruling out a KPZ
term.

\begin{figure}
\begin{center}
\includegraphics{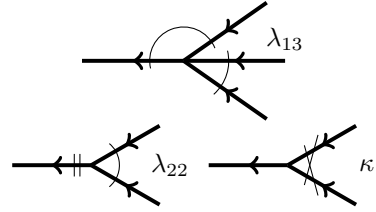}
\end{center}
\caption{\flabel{coupling_feynman_diagrams}
The vertices of the three non-linearities in \Eref{full_theory}. Thick lines
with arrows denote the bare propagator $(-\imag \omega + \nu_2 \kvec^2 +
\nu_4 (\kvec^2)^2)^{-1}$. Arcs and dashes drawn in narrow lines denote
inner products of factors of $\kvec$. The vertex corresponding to
$\kappa$ has all factors of $\kvec$ on one side, in the form
$(\kvec_1\times\kvec_2)^2=\kvec_1^2\kvec_2^2-(\kvec_1\cdot\kvec_2)^2$.
}
\end{figure}

Re-writing two of the couplings in a computationally convenient
form using $\tilde{\lambda}_{22}=\lambda_{22}-\kappa/2$ and
$\tilde{\kappa}=\kappa$:
\begin{multline}\elabel{rewrite_for_kappa}
\underbrace{(\lambda_{22}-\frac{\kappa}{2})}_{\tilde{\lambda}_{22}} \nabla^{2} \Big(\nabla\phi\Big)^{2}+
\underbrace{\kappa}_{\tilde{\kappa}} \nabla\left(\nabla\phi\nabla^{2}\phi\right)\\
=\lambda_{22}\nabla^{2}\Big(\nabla\phi\Big)^{2}+\kappa\left[\nabla\left(\nabla\phi\nabla^{2}\phi\right)-\frac{1}{2}\nabla^{2}\Big(\nabla\phi\Big)^{2}\right]
\end{multline}
Giving rise to the vertices shown in \Fref{coupling_feynman_diagrams}.
The original formulation of this Langevin equation
\cite{Villain91,LaiDasSarmaPRL91} describing epitaxial growth on a $d$
dimensional substrate did not contain the $\kappa$ term
\Eref{rewrite_for_kappa}. However it should appear in two ways. Either
it should be there from the start from the same symmetry and
conservation arguments that produce $\lambda_{22}$. Or if it is absent
it is generated anyway from $\lambda_{13}$ and $\lambda_{22}$ under
renormalisation of the full theory \Eref{full_theory}. It is this
$\kappa$ non-linearity that provides a natural mound forming mechanism.
The $\kappa$ notation has be used in keeping with the term's appearance
in an unrelated restricted solid-on-solid model \cite{LazeridesPRE06}.

The original VLDS formulation was derived by considering the most
general fourth order equation consistent with the symmetries of the
problem, with further refinements coming from other physical insights.
Later a second way of deriving the equation was sought by considering
the transition rules on a lattice model of the surface
\cite{VvPRE93,HuangGuPRE96,VvHasPRL07,VvHasPRE07}, finding ways to turn
them into a continuum equation and dropping terms considered to be
irrelevant for the scaling behaviour (seeking to capture the
``qualitative features of the surface morphology'' \cite{VvHasPRE07}).
Virtually all of the analysis of the continuum equations has been done
using the Dynamic Renormalisation Group (DRG), a technique that is an
extension of Wilson's approach to Renormalisation Group (RG)
calculations, in that a cutoff is used and momentum shells are
integrated out to find the differential flow equations of the couplings
\citecommented{\eg}{barabasistanley}.  Instead of using DRG one can
instead turn the Langevin equation in question into a field theory, the
programme pursued below. This entails forming an action constrained by
the Langevin equation in order to produce a (moment) generating
functional. This approach was developed independently by Janssen
\cite{janssenpaper76} and De Dominicis \cite{dedominicispaper78}.

Prosecuting a renormalised field theory of \Eref{full_theory} to one
loop various theories of epitaxial growth are recovered as fixed points
or limiting cases. In all dimensions \Eref{full_theory} displays
Edwards-Wilkinson (EW) behaviour for $\nu_{2}$ greater than the critical
point $\nu_{2}^{c}$, with $\nu_{2}^{c}=0$  in the absence of
non-linearities. For $\nu_{2}<\nu_{2}^{c}$ the bare propagator acquires
a pole at the characteristic wavelength $\sqrt{\nu_{4}/\nu_{2}}$. In
dimensions above the critical dimension, $d_c=4$,  \Eref{full_theory} is
trivially governed by Mullins-Herring (MH) behaviour at critical
$\nu_2=\nu_2^c$. However, in dimensions $d=d_c-\epsilon<d_c$, the
non-linearities $\lambda_{13}$, $\lambda_{22}$ and $\kappa$ are all
(equally) relevant and produce the non-trivial scaling behaviour
characterised below. In the presence of $\lambda_{13}$, the critical
$\nu_2^c$ becomes negative and there is no longer a generic mechanism
guaranteeing $\nu_2=\nu_2^c$ as exists in the case $\nu_2^c=0$, \ie the
non-trivial scaling is visible only after tuning to a critical point.
This is the reason why $\lambda_{13}$ has been dismissed originally.
However, as pointed out by Haselwandter and Vvedensky \cite{VvHasPRL07},
the non-trivial behaviour of the full theory might still be visible on
an intermediate scale, beyond which EW might rule. In
particular, given that $\kappa$ is generated under renormalisation,
finite mounding is expected to occur generically. What is normally
referred to as the VLDS equation, however, is \Eref{full_theory} with
$\nu_{2}$, $\lambda_{13}$, $\kappa$ set to zero.

The ultraviolet is regularised in a perturbation theory in small
$\epsilon=d_c-d>0$, where the reparamaterised, dimensionless couplings read:
\begin{equation}\elabel{def_couplings}
g=\frac{\Gamma^{2}}{(4\pi)^{2}}\frac{\lambda_{13}}{\nu_{4}^{2-\epsilon/2}}
\qquad
\lambda=\frac{\lambda_{22}^{2}}{\nu_{4}\lambda_{13}}
\qquad
\chi=\frac{\kappa}{\lambda_{22}} 
\end{equation}
The renormalisation of the couplings is determined by accounting for all
logarithmically  divergent diagrams contributing to the proper vertices
$\Gamma^{(1,1)}$, $\Gamma^{(1,2)}$ and $\Gamma^{(1,3)}$, the functional
derivatives of the Legendre transform $\Gamma$ of the cumulant generating function, \ie:
\begin{multline*}
\Gamma^{(n,m)} \left(\kvec_1,\dots,\omega_{n+m}; \nu_2, \nu_4, \Gamma^2, \lambda_{13}, \lambda_{22}, \kappa\right) = \\
\prod_{i=1}^{n} 
  \frac{\delta}{\delta \phitilde(\kvec_i,\omega_i)}
\prod_{j=1}^{m}
  \frac{\delta}{\delta \phi(\kvec_{n+j}, \omega_{n+j}) }\\
    \Gamma\left([\phi], [\phitilde]; \nu_2, \nu_4, \Gamma^2,
    \lambda_{13}, \lambda_{22}, \kappa\right),
\end{multline*}
as derived from the one-particle irreducible, amputated diagrams
contributing to the corresponding correlation function. They give rise
to the renormalisation of the couplings $\alpha$ in the form
$\alpha^R=Z_{\alpha}\alpha$. The infrared, on the other hand, is
regularised by the mass $\nu_2\ne0$, with $\nu^R_2=Z_2 \nu_2 \mu^{-2}$,
renormalisation point $\nu^R_2=1$ and arbitrary inverse length $\mu$.

The infrared stable fixed point is found as a root of the set of
beta-functions, $\beta_{a}=\plaind\ln{\alpha} / \plaind \ln{\mu}
|_{\text{bare}}$, where the derivative is to be taken for every coupling
$\alpha$ at constant bare couplings: 
\begin{eqnarray*}
\tilde{\beta}_{g}&=&-\epsilon+\left(5-(2-\epsilon/2)\lambda\left[\frac{5}{2}\chi^{2}-3\chi-1\right]\right)g\nonumber\\
\tilde{\beta}_{\lambda}&=&\left(4-\lambda\left[\frac{5}{2}\chi^{2}-3\chi-1\right]\right)g\\
\tilde{\beta}_{\chi}&=&\left(\frac{1}{2}+\frac{1}{\chi}\right)g\nonumber,
\end{eqnarray*}
where we define $\tilde{\beta}_{\alpha}=\beta_{\alpha}/\alpha$ for convenience. 
The Wilson gamma-functions are correspondingly defined as $\gamma_{\alpha}=\plaind\ln{Z_{a}}/\plaind\ln{\mu}|_{\text{bare}}$:
\begin{eqnarray*}
&&\gamma_{2}=3g,\qquad\gamma_{4}=\lambda\left(\frac{5}{2}\chi^{2}-3\chi-1\right)g,\\
&&\gamma_{g}=\left(5-(2-\epsilon/2)\lambda\left[\frac{5}{2}\chi^{2}-3\chi-1\right]\right)g,\\
&&\gamma_{\lambda}=\left(4-\lambda\left[\frac{5}{2}\chi^{2}-3\chi-1\right]\right)g,
\quad \gamma_{\chi}=\left(\frac{1}{2}+\frac{1}{\chi}\right)g,\\
&&\gamma_{22}=\frac{9}{2}g,\quad\gamma_{\kappa}=\left(5+\frac{1}{\chi}\right)g,\quad\gamma_{13}=5g
\end{eqnarray*}
As the noise does not renormalise at any order, $\gamma_{\Gamma}=0$.

The scaling behaviour of the proper vertices,
\begin{multline}\elabel{full_vertex_coupling_scaling}
\hat{\Gamma}^{(n,m)}\left(\kvec, \omega;
\nu_{2},\nu_{4},\Gamma^2,\lambda_{13}, \lambda_{22}, \kappa\right)\\
=l^{-\frac{m}{2}(d-4)-\frac{n}{2}(d+4) + d+4 -\gamma_4(\half(m-n)+1}\\
\hat{\Gamma}^{(n,m)}\left(\frac{\kvec}{l}, \frac{\omega}{l^{\gamma_{4}+4}};
\nu_{2}l^{\gamma_{2}-\gamma_{4}-2},
\nu_{4}, \Gamma^2,
\lambda_{13}l^{-2\gamma_{4}-\epsilon+\gamma_{13}},\right.\\
\left.\lambda_{22}l^{\gamma_{22}-\frac{3\gamma_{4}}{2}-\frac{\epsilon}{2}},
\kappa l^{\gamma_{\kappa}-\frac{3\gamma_{4}}{2}-\frac{\epsilon}{2}}\right)
\end{multline}
determines the exponents natural to the field theory:
\begin{multline}
\hat{\Gamma}^{(n,m)}\left(\kvec, \omega;
\nu_{2},\nu_{4},\Gamma^2,\lambda_{13}, \lambda_{22}, \kappa\right)\\
=l^{-\frac{m}{2}(d-4+\eta)-\frac{n}{2}(d+4+\etatilde) + d+4 +\delta}\\
\hat{\Gamma}^{(n,m)}\left(\frac{\kvec}{l}, \frac{\omega}{l^{z}};
\frac{\nu_{2}}{l^{1/\nu}},
\nu_{4}, \Gamma^2,
\frac{\lambda_{13}}{l^{\pi_{13}}},
\frac{\lambda_{22}}{l^{\pi_{22}}},
\frac{\kappa}{l^{\pi_{\kappa}}}\right) \ ,
\end{multline}
Where the hat, $\hat{\cdot}$, indicates that the Dirac $\delta$ function
from momentum conservation by translational invariance has been divided
out, $\Gamma^{(n,m)}=\delta(\kvec_1+\ldots+\kvec_{n+m})
\delta(\omega_1+\ldots+\omega_{n+m}) \hat{\Gamma}^{(n,m)}$ 
eliminating one pair of arguments $\kvec$, $\omega$. Normally, growth
exponents ($\alpha$, $z$ and $\beta=\alpha/z$) characterise the approach
of stationarity from a flat initial configuration and the finite size
scaling of the roughness. In a field theory this is not particularly
germane, so the exponents are equivalently \citecommented{but
see}{Lopez:1999} defined on the basis of the two point correlation
function:
\begin{eqnarray*}
&&\hat{\mathbf{G}}^{20}(\qvec,\omega)= a
|\qvec|^{-(d+z+2\alpha)}\;\mathcal{G}\left(\frac{\omega}{b |\qvec|^{z}}\right)
= - \hat{\Gamma}^{20} (\qvec,\omega)\\
&\times &\left|\hat{\Gamma}^{11}(\qvec,\omega)\right|^{-2}= 2 \Gamma^2
\left||\qvec|^{\gamma_{4}+4}\;\hat{\Gamma}^{11}\left(\frac{\qvec}{|\qvec|},\frac{\omega}{b
|\qvec|^{\gamma_{4}+4}}\right)\right|^{-2}
\end{eqnarray*}
With suitable metric factor $a$ and $b$ and universal scaling function
$\mathcal{G}(x)$. In the following, we therefore focus on the exponents
\begin{equation}
\delta=-\gamma_{4},\quad 
\nu=\frac{1}{\gamma_{4}+2-\gamma_{2}},\quad
z=\gamma_{4}+4,\quad
\alpha=\frac{\epsilon+\gamma_4}{2}.
\end{equation}

The simultaneous roots of the beta-functions give the fixed points of the theory. The infrared stable one at 
\begin{equation}\elabel{infra_stable_fp}
\chi=-2, \qquad \lambda=0\ \text{ and }\ g=\epsilon/5 
\end{equation}
gives $\gamma_4=0$ and $\gamma_2=3\epsilon/5$ and thus
\begin{equation}\elabel{full_expos}
\delta=0, \qquad
\nu=\frac{1}{2}+\frac{3\epsilon}{20},\qquad
z=4,\qquad
\alpha=\frac{\epsilon}{2}, 
\end{equation}
are the exponents of the full VLDS equation \Eref{full_theory} at
the critical point $\nu_2=\nu_2^c$. These are the same $\alpha$ and $z$ exponents as for
MH. With $\lambda=0$ this implies both $\lambda_{22}$ and
$\kappa$ are zero at the fixed point. The renormalisation of the full
theory \Eref{full_theory} is driven by $\lambda_{13}$. In agreement with
previous results \cite{LaiDasSarmaPRL91} with only $\lambda_{22}$ we find that $\lambda_{22}$ and $\kappa$
do not renormalise themselves at one loop. The contributions from diagrams involving only these two
couplings neatly cancel at one loop but not at higher orders
\cite{JanssenPRL97}. While $\kappa$ and $\lambda_{22}$ do not
generate each other, they do mix under renormalisation.

Of the two unstable fixed points, the trivial one, $g=0$, deserves
further attention. As $g=0$ implies $\lambda_{13}=0$ which causes
problems with the definition of $\lambda$, \Eref{def_couplings}, it is
not legitimate to naively read all the gamma-functions as zero and
extract exact MH behaviour. The scaling behaviour of this trivial
fixed point is normally referred to as the VLDS fixed point, observed
either by taking the limit $\lambda_{13}\rightarrow0$ or by removing it
from the initial Langevin equation \cite{LaiDasSarmaPRL91,VvHasPRE07},
possible as $\lambda_{13}$ is not generated by $\lambda_{22}$ or $\kappa$. Keeping $\chi$ and replacing $\lambda$ by
\begin{equation}\elabel{def_psi}
\psi=\frac{\Gamma^{2}}{(4\pi)^{2}}\frac{\lambda_{22}^{2}}{\nu_{4}^{3-\epsilon/2}}
\end{equation}
gives $\tilde{\beta}_\psi=-\epsilon-(3-\epsilon/2)[(5/2)\chi^{2}-3\chi-1]\psi$ and
\begin{equation}\elabel{VLDS_expos}
\delta=\frac{\epsilon}{3}, \qquad 
\nu=\frac{1}{2}+\frac{\epsilon}{12}, \qquad
z=4-\frac{\epsilon}{3}, \qquad 
\alpha=\frac{\epsilon}{3},
\end{equation}
where $z$ and $\alpha$ are the traditional one loop VLDS exponents
\cite{LaiDasSarmaPRL91}. If only the $\lambda_{22}$ non-linearity is present, that is $\chi=0$ is
taken at the trivial fixed point, then \Eref{full_theory} becomes the
original VLDS formulation. Interestingly
the corrections to scaling at two-loops predicted by
Janssen \cite{JanssenPRL97} on the basis of $\lambda_{22}$ alone were
found to be too small in numerical lattice
simulations \cite{YookKimKimPRE97,YookLeeKimPRE98,ReisPRE04}, this may be
due to a $\kappa$ correction. Renormalisation does however impose bounds
on $\chi$, or the fixed point of $\psi$ would oblige an unphysical
$\nu_{4}$. To one loop, in order to get sensible results $(5/2)\chi^{2}-3\chi-1$
needs to be negative, $\chi\in
[\frac{3-\sqrt{19}}{5},\frac{3+\sqrt{19}}{5}]$, sufficiently large $\kappa$ violates this. 
However the same exponents emerge \cite{EscuderoKorutcheva:2012} if
implemented without  consideration of this.

We observe that the $\kappa$ coupling on its own is equivalent to a
model proposed by Escudero \cite{EscuderoPRL08} to reproduce VLDS
behaviour. The non-linearity proposed in two dimensions,
$\partial_{xx}\phi\;\partial_{yy}\phi-(\partial_{xy}\phi)^{2}$, is
exactly the vertex parameterised by $\kappa$ in
\Eref{rewrite_for_kappa}. Pursuing the calculation with only $\kappa$ is
especially straight forward, power counting reveals that diagrams for
its renormalisation constructed solely from the $\kappa$ and the noise
vertices are always ultraviolet finite; in the absence of any
other coupling $\kappa$ is not renormalised at any order. The
only renormalisation is of the propagator, with one diagram at
one-loop order. Hence scaling laws are not corrected to any order, yet
Janssen's general insight is not wrong \cite{JanssenPRL97}; it is a
peculiarity of having only $\kappa$ that leads to non-renormalisation of
the coupling, as opposed to a neat cancellation (only) at one-loop when
$\lambda_{22}$ is also present. Perhaps other analyses finding exact 
scaling laws have inadvertently examined this case rather than the VLDS equation. 
It is now quite apparent why this model
reproduces VLDS exponents. It is now also apparent that the model's
infrared stable fixed point is unphysical (to one loop), and its
behaviour thus not assessable by perturbation theory.

As has been observed from its two-dimensional form \cite{EscuderoPRL08},
the $\kappa$ non-linearity favours mound formation. In addition to
Ehrlich-Schwoebel (ES) barriers expressed through $\nu_{2}$
\cite{Krug:1997,KrugMichely2004} it provides a natural
mechanism at the level of the continuum equations for mound formation.
In the presence of $\lambda_{13}$, at the critical point, mounding is
suppressed on the large scale, \Eref{infra_stable_fp}, yet visible at
and below (transient) length scales $\propto \kappa^{5/(2\epsilon)}$,
\Eref{full_vertex_coupling_scaling}. Appealing to recent work done on
coupling flow in models of ideal MBE \cite{VvHasPRL07} provides some
theoretical justification for transient observation of mounding. At the
trivial fixed point, on the other hand, VLDS scaling applies generically
as $\nu_2^c$ does not suffer an additive renormalisation, and mounding
may be present on all length scales.

A connection may also be made to the dynamics used in lattice models of
ideal MBE. There have been several investigations into the link between
rules of movement in lattice models to the terms in continuum
equations that represent them, usually concentrating on one-dimension
\cite{DasSarmaNumPRE96}. Hugston and Ketterl \cite{HugstonPRE99} showed
that going from lattice rules that seem intuitive, or even
computationally convenient, to continuum equations is subtle and fraught
with unintended consequences. Step edge diffusion \cite{DasSarmaPRB01},
appearing in two-dimensions, has been proffered as an additional
mechanism to ES barriers that leads to unstable mounding, for example in
the two-dimensional WV model \cite{DasSarmaPRE02}. The lattice rules for 
the DT model are slightly different and result in EW behaviour instead of 
mounding in two-dimensions. Differences in lattice rules may 
well be the distinction between having
$\kappa$ or not in the continuum equation for a lattice model.

Observing that the terms parameterised by $\lambda_{22}$ and $\kappa$
are equivalent in one-dimension offers an explanation as to why $\kappa$
was absent in the original derivation of the VLDS equation. It
was not generated by renormalisation since the analysis
immediately focused on $\lambda_{13}=0$ to prevent $\nu_2$ generation.
If ever $\lambda_{13}\neq0$ then $\lambda_{22}$ would be set to
zero ``without loss of generality'', as $\lambda_{13}$ was deemed ``more
relevant'' \cite{DasSarma2loopPRE94}, further analysis halted by the
generation of $\nu_{2}$. One analysis \cite{VvHasPRL07,VvHasPRE07} did not commit
this omission, but unfortunately missed the diagram generating $\kappa$. 
A similar commentary applies to the use of master equations to
generate continuum equations from lattice rules. The most telling sign
is seen when, in developing a continuum equation for the WV model,
derivatives are rearranged in one-dimension and the result carried over
to higher dimensions where, however, it is invalid \citecommented{\eg
Eq.~(23) from Eq.~(21) in}{HuangGuPRE96}. One might wonder whether basic
errors in multivariate calculus \cite{Stephenson}, such as assuming
wrongly that $\nabla(\nabla{\phi})^{3}$ equals
$3(\nabla^{2}\phi)(\nabla{\phi})^{2}$ \cite{DasSarmaNumPRE96} have thus
far concealed the $\kappa$ term. Fourier transforming such terms
immediately reveals the problem \cite{Stephenson}.

In summary we have shown that the original VLDS formulation generates a
mounding term, $\kappa$, overlooked in previous studies, partly due to
non-linearities being deliberately omitted, partly
the renormalisation schemes employed missed the generation of
$\kappa$ and partly from an apparent misunderstanding of
basic multivariate calculus. This $\kappa$ term might effectively capture certain lattice rules that give
rise to mounding in computer simulations; inadvertently studying
models with only $\kappa$ present may lead to concluding scaling relations are in general exact,
whereas the presence of $\kappa$ with $\lambda_{22}$ may account for differences in
expected scaling corrections.

\bibliography{pr_ideal_mbe}

\begin{thebibliography}{29}
\expandafter\ifx\csname natexlab\endcsname\relax\def\natexlab#1{#1}\fi
\expandafter\ifx\csname bibnamefont\endcsname\relax
  \def\bibnamefont#1{#1}\fi
\expandafter\ifx\csname bibfnamefont\endcsname\relax
  \def\bibfnamefont#1{#1}\fi
\expandafter\ifx\csname citenamefont\endcsname\relax
  \def\citenamefont#1{#1}\fi
\expandafter\ifx\csname url\endcsname\relax
  \def\url#1{\texttt{#1}}\fi
\expandafter\ifx\csname urlprefix\endcsname\relax\def\urlprefix{URL }\fi
\providecommand{\bibinfo}[2]{#2}
\providecommand{\eprint}[2][]{\url{#2}}

\bibitem[{\citenamefont{Barabasi and Stanley}(1995)}]{barabasistanley}
\bibinfo{author}{\bibfnamefont{A.~L.} \bibnamefont{Barabasi}} \bibnamefont{and}
  \bibinfo{author}{\bibfnamefont{H.~E.} \bibnamefont{Stanley}},
  \emph{\bibinfo{title}{{Fractal Concepts in Surface Growth}}}
  (\bibinfo{publisher}{Cambridge University Press}, \bibinfo{year}{1995}).

\bibitem[{\citenamefont{Pimpinelli and Villain}(1998)}]{pimpinellicrystals}
\bibinfo{author}{\bibfnamefont{A.}~\bibnamefont{Pimpinelli}} \bibnamefont{and}
  \bibinfo{author}{\bibfnamefont{J.}~\bibnamefont{Villain}},
  \emph{\bibinfo{title}{Physics of crystal growth}}
  (\bibinfo{publisher}{Cambridge University Press}, \bibinfo{year}{1998}).

\bibitem[{\citenamefont{Krug}(1997)}]{Krug:1997}
\bibinfo{author}{\bibfnamefont{J.}~\bibnamefont{Krug}}, \bibinfo{journal}{Adv.
  Phys.} \textbf{\bibinfo{volume}{46}}, \bibinfo{pages}{139}
  (\bibinfo{year}{1997}).

\bibitem[{\citenamefont{Michely and Krug}(2004)}]{KrugMichely2004}
\bibinfo{author}{\bibfnamefont{T.}~\bibnamefont{Michely}} \bibnamefont{and}
  \bibinfo{author}{\bibfnamefont{J.}~\bibnamefont{Krug}},
  \emph{\bibinfo{title}{Islands, Mounds, and Atoms}}, Springer series in
  surface sciences (\bibinfo{publisher}{Springer}, \bibinfo{year}{2004}).

\bibitem[{\citenamefont{Kardar et~al.}(1986)\citenamefont{Kardar, Parisi, and
  Zhang}}]{kpz}
\bibinfo{author}{\bibfnamefont{M.}~\bibnamefont{Kardar}},
  \bibinfo{author}{\bibfnamefont{G.}~\bibnamefont{Parisi}}, \bibnamefont{and}
  \bibinfo{author}{\bibfnamefont{Y.-C.} \bibnamefont{Zhang}},
  \bibinfo{journal}{Phys. Rev. Lett.} \textbf{\bibinfo{volume}{56}},
  \bibinfo{pages}{889} (\bibinfo{year}{1986}).

\bibitem[{\citenamefont{Wolf and Villain}(1990)}]{wolfvillain90}
\bibinfo{author}{\bibfnamefont{D.~E.} \bibnamefont{Wolf}} \bibnamefont{and}
  \bibinfo{author}{\bibfnamefont{J.}~\bibnamefont{Villain}},
  \bibinfo{journal}{Europhys. Lett.} \textbf{\bibinfo{volume}{13}},
  \bibinfo{pages}{389} (\bibinfo{year}{1990}).

\bibitem[{\citenamefont{Das~Sarma and Tamborenea}(1991)}]{DTmodelPRL91}
\bibinfo{author}{\bibfnamefont{S.}~\bibnamefont{Das~Sarma}} \bibnamefont{and}
  \bibinfo{author}{\bibfnamefont{P.}~\bibnamefont{Tamborenea}},
  \bibinfo{journal}{Phys. Rev. Lett.} \textbf{\bibinfo{volume}{66}},
  \bibinfo{pages}{325} (\bibinfo{year}{1991}).

\bibitem[{\citenamefont{Villain}(1991)}]{Villain91}
\bibinfo{author}{\bibfnamefont{J.}~\bibnamefont{Villain}}, \bibinfo{journal}{J.
  Phys. I (France)} \textbf{\bibinfo{volume}{1}}, \bibinfo{pages}{19}
  (\bibinfo{year}{1991}).

\bibitem[{\citenamefont{Lai and Das~Sarma}(1991)}]{LaiDasSarmaPRL91}
\bibinfo{author}{\bibfnamefont{Z.-W.} \bibnamefont{Lai}} \bibnamefont{and}
  \bibinfo{author}{\bibfnamefont{S.}~\bibnamefont{Das~Sarma}},
  \bibinfo{journal}{Phys. Rev. Lett.} \textbf{\bibinfo{volume}{66}},
  \bibinfo{pages}{2348} (\bibinfo{year}{1991}).

\bibitem[{\citenamefont{Lazarides}(2006)}]{LazeridesPRE06}
\bibinfo{author}{\bibfnamefont{A.}~\bibnamefont{Lazarides}},
  \bibinfo{journal}{Phys. Rev. E} \textbf{\bibinfo{volume}{73}},
  \bibinfo{pages}{041605} (\bibinfo{year}{2006}).

\bibitem[{\citenamefont{Vvedensky et~al.}(1993)\citenamefont{Vvedensky,
  Zangwill, Luse, and Wilby}}]{VvPRE93}
\bibinfo{author}{\bibfnamefont{D.~D.} \bibnamefont{Vvedensky}},
  \bibinfo{author}{\bibfnamefont{A.}~\bibnamefont{Zangwill}},
  \bibinfo{author}{\bibfnamefont{C.~N.} \bibnamefont{Luse}}, \bibnamefont{and}
  \bibinfo{author}{\bibfnamefont{M.~R.} \bibnamefont{Wilby}},
  \bibinfo{journal}{Phys. Rev. E} \textbf{\bibinfo{volume}{48}},
  \bibinfo{pages}{852} (\bibinfo{year}{1993}).

\bibitem[{\citenamefont{Huang and Gu}(1996)}]{HuangGuPRE96}
\bibinfo{author}{\bibfnamefont{Z.-F.} \bibnamefont{Huang}} \bibnamefont{and}
  \bibinfo{author}{\bibfnamefont{B.-L.} \bibnamefont{Gu}},
  \bibinfo{journal}{Phys. Rev. E} \textbf{\bibinfo{volume}{54}},
  \bibinfo{pages}{5935} (\bibinfo{year}{1996}).

\bibitem[{\citenamefont{Haselwandter and
  Vvedensky}(2007{\natexlab{a}})}]{VvHasPRL07}
\bibinfo{author}{\bibfnamefont{C.~A.} \bibnamefont{Haselwandter}}
  \bibnamefont{and} \bibinfo{author}{\bibfnamefont{D.~D.}
  \bibnamefont{Vvedensky}}, \bibinfo{journal}{Phys. Rev. Lett.}
  \textbf{\bibinfo{volume}{98}}, \bibinfo{pages}{046102}
  (\bibinfo{year}{2007}{\natexlab{a}}).

\bibitem[{\citenamefont{Haselwandter and
  Vvedensky}(2007{\natexlab{b}})}]{VvHasPRE07}
\bibinfo{author}{\bibfnamefont{C.~A.} \bibnamefont{Haselwandter}}
  \bibnamefont{and} \bibinfo{author}{\bibfnamefont{D.~D.}
  \bibnamefont{Vvedensky}}, \bibinfo{journal}{Phys. Rev. E}
  \textbf{\bibinfo{volume}{76}}, \bibinfo{pages}{041115}
  (\bibinfo{year}{2007}{\natexlab{b}}).

\bibitem[{\citenamefont{Janssen}(1976)}]{janssenpaper76}
\bibinfo{author}{\bibfnamefont{H.-K.} \bibnamefont{Janssen}},
  \bibinfo{journal}{Z. Phys. B} \textbf{\bibinfo{volume}{23}},
  \bibinfo{pages}{377} (\bibinfo{year}{1976}).

\bibitem[{\citenamefont{De~Dominicis and Peliti}(1978)}]{dedominicispaper78}
\bibinfo{author}{\bibfnamefont{C.}~\bibnamefont{De~Dominicis}}
  \bibnamefont{and} \bibinfo{author}{\bibfnamefont{L.}~\bibnamefont{Peliti}},
  \bibinfo{journal}{Phys. Rev. B} \textbf{\bibinfo{volume}{18}},
  \bibinfo{pages}{353} (\bibinfo{year}{1978}).

\bibitem[{\citenamefont{L{\'o}pez}(1999)}]{Lopez:1999}
\bibinfo{author}{\bibfnamefont{J.~M.} \bibnamefont{L{\'o}pez}},
  \bibinfo{journal}{Phys. Rev. Lett.} \textbf{\bibinfo{volume}{83}},
  \bibinfo{pages}{4594} (\bibinfo{year}{1999}).

\bibitem[{\citenamefont{Janssen}(1997)}]{JanssenPRL97}
\bibinfo{author}{\bibfnamefont{H.~K.} \bibnamefont{Janssen}},
  \bibinfo{journal}{Phys. Rev. Lett.} \textbf{\bibinfo{volume}{78}},
  \bibinfo{pages}{1082} (\bibinfo{year}{1997}).

\bibitem[{\citenamefont{Yook et~al.}(1997)\citenamefont{Yook, Kim, and
  Kim}}]{YookKimKimPRE97}
\bibinfo{author}{\bibfnamefont{S.~H.} \bibnamefont{Yook}},
  \bibinfo{author}{\bibfnamefont{J.~M.} \bibnamefont{Kim}}, \bibnamefont{and}
  \bibinfo{author}{\bibfnamefont{Y.}~\bibnamefont{Kim}},
  \bibinfo{journal}{Phys. Rev. E} \textbf{\bibinfo{volume}{56}},
  \bibinfo{pages}{4085} (\bibinfo{year}{1997}).

\bibitem[{\citenamefont{Yook et~al.}(1998)\citenamefont{Yook, Lee, and
  Kim}}]{YookLeeKimPRE98}
\bibinfo{author}{\bibfnamefont{S.~H.} \bibnamefont{Yook}},
  \bibinfo{author}{\bibfnamefont{C.~K.} \bibnamefont{Lee}}, \bibnamefont{and}
  \bibinfo{author}{\bibfnamefont{Y.}~\bibnamefont{Kim}},
  \bibinfo{journal}{Phys. Rev. E} \textbf{\bibinfo{volume}{58}},
  \bibinfo{pages}{5150} (\bibinfo{year}{1998}).

\bibitem[{\citenamefont{Aar\~ao Reis}(2004)}]{ReisPRE04}
\bibinfo{author}{\bibfnamefont{F.~D.~A.} \bibnamefont{Aar\~ao Reis}},
  \bibinfo{journal}{Phys. Rev. E} \textbf{\bibinfo{volume}{70}},
  \bibinfo{pages}{031607} (\bibinfo{year}{2004}).

\bibitem[{\citenamefont{Escudero and
  Korutcheva}(2012)}]{EscuderoKorutcheva:2012}
\bibinfo{author}{\bibfnamefont{C.}~\bibnamefont{Escudero}} \bibnamefont{and}
  \bibinfo{author}{\bibfnamefont{E.}~\bibnamefont{Korutcheva}},
  \bibinfo{journal}{J. Phys. A: Math. Theor.} \textbf{\bibinfo{volume}{45}},
  \bibinfo{pages}{125005} (\bibinfo{year}{2012}).

\bibitem[{\citenamefont{Escudero}(2008)}]{EscuderoPRL08}
\bibinfo{author}{\bibfnamefont{C.}~\bibnamefont{Escudero}},
  \bibinfo{journal}{Phys. Rev. Lett.} \textbf{\bibinfo{volume}{101}},
  \bibinfo{pages}{196102} (\bibinfo{year}{2008}).

\bibitem[{\citenamefont{Das~Sarma et~al.}(1996)\citenamefont{Das~Sarma,
  Lanczycki, Kotlyar, and Ghaisas}}]{DasSarmaNumPRE96}
\bibinfo{author}{\bibfnamefont{S.}~\bibnamefont{Das~Sarma}},
  \bibinfo{author}{\bibfnamefont{C.~J.} \bibnamefont{Lanczycki}},
  \bibinfo{author}{\bibfnamefont{R.}~\bibnamefont{Kotlyar}}, \bibnamefont{and}
  \bibinfo{author}{\bibfnamefont{S.~V.} \bibnamefont{Ghaisas}},
  \bibinfo{journal}{Phys. Rev. E} \textbf{\bibinfo{volume}{53}},
  \bibinfo{pages}{359} (\bibinfo{year}{1996}).

\bibitem[{\citenamefont{Hagston and Ketterl}(1999)}]{HugstonPRE99}
\bibinfo{author}{\bibfnamefont{W.~E.} \bibnamefont{Hagston}} \bibnamefont{and}
  \bibinfo{author}{\bibfnamefont{H.}~\bibnamefont{Ketterl}},
  \bibinfo{journal}{Phys. Rev. E} \textbf{\bibinfo{volume}{59}},
  \bibinfo{pages}{2699} (\bibinfo{year}{1999}).

\bibitem[{\citenamefont{Chatraphorn et~al.}(2001)\citenamefont{Chatraphorn,
  Toroczkai, and Das~Sarma}}]{DasSarmaPRB01}
\bibinfo{author}{\bibfnamefont{P.~P.} \bibnamefont{Chatraphorn}},
  \bibinfo{author}{\bibfnamefont{Z.}~\bibnamefont{Toroczkai}},
  \bibnamefont{and}
  \bibinfo{author}{\bibfnamefont{S.}~\bibnamefont{Das~Sarma}},
  \bibinfo{journal}{Phys. Rev. B} \textbf{\bibinfo{volume}{64}},
  \bibinfo{pages}{205407} (\bibinfo{year}{2001}).

\bibitem[{\citenamefont{Das~Sarma et~al.}(2002)\citenamefont{Das~Sarma,
  Chatraphorn, and Toroczkai}}]{DasSarmaPRE02}
\bibinfo{author}{\bibfnamefont{S.}~\bibnamefont{Das~Sarma}},
  \bibinfo{author}{\bibfnamefont{P.~P.} \bibnamefont{Chatraphorn}},
  \bibnamefont{and}
  \bibinfo{author}{\bibfnamefont{Z.}~\bibnamefont{Toroczkai}},
  \bibinfo{journal}{Phys. Rev. E} \textbf{\bibinfo{volume}{65}},
  \bibinfo{pages}{036144} (\bibinfo{year}{2002}).

\bibitem[{\citenamefont{Das~Sarma and Kotlyar}(1994)}]{DasSarma2loopPRE94}
\bibinfo{author}{\bibfnamefont{S.}~\bibnamefont{Das~Sarma}} \bibnamefont{and}
  \bibinfo{author}{\bibfnamefont{R.}~\bibnamefont{Kotlyar}},
  \bibinfo{journal}{Phys. Rev. E} \textbf{\bibinfo{volume}{50}},
  \bibinfo{pages}{R4275} (\bibinfo{year}{1994}).

\bibitem[{\citenamefont{Stephenson}(1973)}]{Stephenson}
\bibinfo{author}{\bibfnamefont{G.}~\bibnamefont{Stephenson}},
  \emph{\bibinfo{title}{Mathematical Methods for Science Students}}
  (\bibinfo{publisher}{Pearson}, \bibinfo{year}{1973}).

\end{thebibliography}

\end{document}